\documentstyle[12pt,aasms4]{article}

\slugcomment{}

\def\kms{km\thinspace s$^{-1}$}
\def\cm2{cm$^{-2}$}
\font \sc = cmr10

\def\HI{H{\sc I}}
\def\HII{H{\sc {II}}}

\def\etal{et al.\ }

\def\eg{e.g.,}

\lefthead{Taylor, Kobulnicky, \& Skillman }
\righthead{CO Emission in Dwarf Galaxies}

\begin{document}

\title{CO Emission in Low Luminosity, HI Rich Galaxies}
\author{Christopher L. Taylor}
\affil{Ruhr-Universit\"at Bochum, Astronomisches Institut \\
Universit\"atsstr 150, D-44870 Bochum \\
Germany}
\authoraddr{Astronomisches Insitut, Universit\"atsstr 150, D-44870 Bochum
Germany \linebreak{\it taylorc@astro.ruhr-uni-bochum.de}}

\author{Henry A. Kobulnicky\footnote{Hubble Fellow}}
\affil{University of California, Santa Cruz \\
Lick Observatory/Board of Studies in Astronomy \\
Santa Cruz, CA, 95064}
\authoraddr{University of California, Santa Cruz, Santa Cruz, CA 95064 
\linebreak {\it chip@ucolick.org}}

\author{Evan D. Skillman}
\affil{University of Minnesota, Department of Astronomy \\
116 Church St. SE \\
Minneapolis, MN 55455}
\authoraddr{Department of Astronomy, 116 Church St. SE Minneapolis, MN 55455
\linebreak {\it skillman@zon.spa.umn.edu}}

\begin{abstract}

We present $^{12}$CO $1 \rightarrow 0$ observations of eleven low
luminosity (M$_B$ $>-18$), \HI--rich dwarf galaxies.  Only the three 
most metal-rich galaxies, with 12+log(O/H)$\approx$ 8.2, are detected.  
Very deep CO spectra of six extremely metal-poor systems
(12+log(O/H)$\le$7.5) yield only low upper limits on the CO surface
brightness, $I_{CO}<0.1$ K \kms.  Three of these six have never before
been observed in a CO line, while the others now have much more
stringent upper limits.  For the very low metallicity galaxy Leo~A, we
do not confirm a previously reported detection in CO, and the limits
are consistent with another recent nondetection.

We combine these new observations with data from the literature to form
a sample of dwarf galaxies which all have CO observations and measured
oxygen abundances.  No known galaxies with 12+log(O/H)$<$7.9
($Z<0.1Z_\odot$) have been detected in CO.  Most of the star-forming
galaxies with higher (12+log(O/H)$>$8.1) metallicities are detected at
similar or higher I$_{CO}$ surface brightnesses.  The data are consistent 
with a strong dependence of the I$_{CO}$/M$_{H_2}~\equiv~X_{CO}$ conversion 
factor on ambient metallicity.  The strikingly low upper limits on some
metal-poor galaxies lead us to predict that the conversion factor is
non-linear, increasing sharply below $\sim$1/10 of the solar
metallicity (12+log(O/H)$\leq$7.9).

\end{abstract}

\section{Introduction}

Carbon monoxide (CO) is commonly used as a tracer of cool molecular
gas, because molecular hydrogen (H$_2$), the dominant species in the
molecular phase, has no strong emission lines from which the column
density of H$_2$ may easily be determined.  Since the rotational
transitions of CO in the millimeter and submillimeter regime are
relatively easy to excite, it is possible to use the luminosity in one
of these lines to estimate the column density and mass of molecular
gas, provided one knows the correct conversion.  The conversion 
factor, $X_{CO}$, from
I$_{CO}$ to N$_{H_2}$ has been determined for the Milky Way galaxy to
be $\sim$ $3~\times~10^{20}$ cm$^{-2}$ (K \kms)$^{-1}$ for the
$^{12}$CO $1 \rightarrow 0$ transition (Strong et al.\markcite{S88}
1988; Scoville \& Sanders\markcite{SS87} 1987).  The application of
this Milky Way value to external galaxies has been controversial, as
the value may depend on the physical conditions in those galaxies which
are difficult to determine observationally and may differ greatly from
those in our own galaxy (Dickman, Snell \& Schloerb\markcite{DSS86}
1986; Israel et al.\ 1986; Maloney \& Black\markcite{MB88} 1988,
hereafter MB88).

One of the characteristics of a galaxy which may affect the relation
between CO luminosity and H$_2$ gas mass is its metal abundance.  If
the abundance of the CO molecule is low, the column density of CO may
not be great enough to allow self shielding from dissociating radiation.
In this case, the size of the CO emitting region within a given
molecular cloud will shrink, while the H$_2$ is unaffected.  Thus the
filling factor will decrease, reducing the CO luminosity for a given
molecular gas mass (MB88).  Rubio, Lequeux \& Boulanger
(1993\markcite{RLB}) have found observational evidence of this effect
in the SMC.  Their data for a number of molecular clouds show a
correlation between cloud size and the CO-to-H$_2$ conversion factor.
They suggest that the smaller clouds are the dense cores of larger
clouds in which the diffuse CO outside the cores has been dissociated
(cf. MB88\markcite{MB88}).

Observational studies have led to conflicting conclusions concerning
the presence of molecular clouds in actively star forming dwarf
galaxies.  Since the pioneering work of Elmegreen, Elmegreen, \& Morris
\markcite{Ee80}(1980), it has been clear that the CO molecule is
difficult to detect in dIs and therefore, that the CO surface
brightnesses are much lower in dIs than in spiral galaxies.  Under the
assumption that the CO/H$_2$ ratio is constant everywhere, this implied
that the molecular gas content of dwarf galaxies must be very low
(Young, Gallagher, \& Hunter \markcite{YGH}1984; Tacconi \& Young
\markcite{TY87}1987).

On the other hand, studies by Wilson (1995,\markcite{W95} hereafter
W95) and Verter \& Hodge (1995\markcite{VH}) have provided strong
evidence that the conversion factor depends on the metal abundance of
the galaxy.  By measuring molecular cloud virial masses and comparing
them to their CO luminosities, W95 \markcite{W95} showed that the
CO-to-H$_2$ conversion increases as the metallicity of the host galaxy
decreases over the range of 8 $\le$ 12 + log(O/H) $\le$ 9.  This
supported the conclusions made by Cohen \etal(1988)\markcite{CDGMRT88} 
and Rubio \etal(1991)\markcite{RGMT91} based on observations of the 
Magellanic Clouds.  Verter \& Hodge (1995)\markcite{VH} added very deep CO 2
$\rightarrow$ 1 observations of the extreme dwarf galaxy GR~8 and and
were unable to detect any CO.  The proximity of GR~8 (2.2 Mpc, Tolstoy
\etal 1995\markcite{TSHM95}) allows very low upper limits on L(CO).   
Combined with an inference of the minimal molecular mass present to 
support the current star formation in GR~8, this was also interpreted as 
an indication of a metallicity dependence of the CO-to-H$_2$ conversion
factor.

Studies of this type have been limited primarily to galaxies of
relatively high metallicities in order to detect CO emission.  Indeed,
the non-detection of GR~8 by Verter \& Hodge (1995)\markcite{VH}, with
an oxygen abundance of 12 + log(O/H) = 7.47 (Skillman, Kennicutt \&
Hodge 1989\markcite{SKH}) illustrates this difficulty.  In fact, the
only low metallicity dwarf irregular galaxy to have been detected in CO
is Leo~A, observed by Tacconi \& Young (1987)\markcite{TY87}, at 12 +
log(O/H) = 7.3.  A recent observation of Leo~A by L. Young (Young 1997,
private communication) failed to confirm the detection of CO in Leo~A.
We decided to try to confirm this important result ourselves, and to
supplement the understanding of CO emission in low metallicity
environments by observing additional metal poor dwarf galaxies.

Here we present $^{12}$CO $1 \rightarrow 0$ observations of 11 galaxies
covering a range of oxygen abundances from 7.3 $\geq$ 12+log(O/H)
$\geq$ 8.2.  Some of these galaxies have been previously observed in
CO.  We confirm previous detections for several galaxies, and we obtain
very deep upper limits for others.  We present the
first published CO data on three galaxies, UGC~4483, DDO~187 and UM422.
Section 2 contains a description of the observations and data reduction
while Section 3 describes the results.  In Section 4 we combine the new
observations with a thorough search of the literature to examine the
relationship between CO surface brightness and metal abundance in these
low-mass systems.

\section{$^{12}$CO Observations and Data Reduction}

\subsection{Observations}

We observed 5 galaxies with the NRAO\footnotemark \footnotetext{The
National Radio Astronomy Observatory is a facility of the National
Science Foundation, operated under cooperative agreement by Associated
Universities, Inc.} 12-m telescope at Kitt Peak, in the $1 \rightarrow
0$ (115 GHz) transition of $^{12}$CO on 5 -- 11 January 1998: Leo~A,
Sextans~A, DDO~210, DDO~187, and Pegasus.  Three galaxies (UM422,
Mrk~178 and UGC~4483) were observed on 10--13 March 1994 and 3 more
(NGC~1569, NGC4214, and NGC5253) on 18--21 June 1995.   The 3 mm SIS
receiver was used with the filterbank spectrometer and a 1 MHz filter,
yielding 256 channels per spectrum, and a channel width of 2.6 \kms.
The receiver was tuned to the central velocity of the \HI\ distribution
in each galaxy.  Operating at 115 GHz, the half power beam width is
55$\arcsec$. System temperatures varied from between $\sim$300 K to 500
K during the course of the observations, infrequently rising higher
during the January 1998 run due to weather conditions.  The pointing
was checked about every two hours by observing Venus or Mars.  The
observations were conducted in beam switching mode, with a beam throw
of 2$\arcmin$ at 1.25 Hz, except for Leo~A and Sextans~A which have too
large an angular extent.  For these galaxies, the absolute position
switching mode was used to ensure a reference beam uncontaminated by
emission from the galaxy.  For comparison, one position in Leo~A was
also observed in beam switching mode.

\subsection{Observing Strategy}

Part of the motivation for our observations was to re-observe Leo~A to
confirm the detection of Tacconi \& Young (1987)\markcite{TY87}.  Since
those original observations, \HI\ interferometer maps for a number of 
dwarf galaxies, including Leo~A, have become available.  In particular,
Young \& Lo (1996)\markcite{YL96} obtained VLA HI observations of Leo~A 
at high spatial and velocity resolution.  They discovered a new, cold 
component to the atomic gas, with a velocity dispersion $\sigma \simeq$ 3.5 
\kms, in addition to the well known warm component with $\sigma \simeq$  9 
\kms.  We reasoned that this cold component of the HI was the most 
likely area to find CO emission in these galaxies, if it was detectable.  
As this cold \HI\ gas is found at the regions of highest HI column density,
we used interferometer maps to direct our CO observations for Leo~A, 
Sextans~A, DDO~210, DDO~187 and Pegasus.  For the other galaxies we pointed
at the center of the stellar distributions, except for NGC~4214, which
we observed at 4 positions within the galaxy.  

\subsection{Data Reduction}

From each individual scan we subtracted  a linear baseline and averaged
multiple scans weighted by a factor of $1/T_{sys}^2$.  We rejected
scans showing unstable baselines which could not be fit with a linear
baseline.  The resulting averaged spectra for the galaxies with the
smallest velocity widths (Leo~A, Sextans~A, DDO~210, DDO~187, and
Pegasus) are smoothed to 5.2, 10.4 and 20.8 \kms\ resolution.   The
remaining galaxies with broader \HI\ line widths we smoothed
to 20.8 \kms.  Total integration times for each object ranged between 6
and 8 hours, including 50\% of the time spent on the sky reference
position.  We searched spectra at each resolution for $^{12}$CO
$1 \rightarrow 0$ emission.  Original resolution and smoothed spectra
for each galaxy appear in Figure~1.  A horizontal bar indicates the
width of the HI profile.  Temperatures indicate brightness units on the
T$_R^*$ scale.

\placefigure{fig1}

\section{Results: CO Detections and Upper Limits}

\subsection{Individual Galaxies}

Table~1 gives the positions, optical diameters, \HI\ heliocentric 
central velocities and velocity widths from the literature for each 
of the newly observed galaxies, as well as the central velocities, 
velocity widths (full widths at 50\% max), the rms noise and the 
integrated intensities (I$_{CO} = \int T^{*}_R dv$) of the CO line 
or the upper limits.  Reported errors on I$_{CO}$ are 
computed from $\sqrt{N}{\times}\sigma_{rms}{\times} \delta v_{chan}$
where $N$ is the number of channels where CO was detected, $\sigma_{rms}$
is the noise in the spectrum, and $\delta v_{chan}$ is the channel~width
in \kms.
\placetable{tab1}

{\it Leo~A---}  Leo~A is a dwarf irregular galaxy for which 
Tacconi \& Young (1987)\markcite{TY87} claim a CO detection.
We observed three positions in this galaxy, two corresponding to the
locations of the cold \HI\ component discovered by Young \& Lo 
(1996)\markcite{YL96}, and the third at the same position observed
by Tacconi \& Young (1987)\markcite{TY87}.  Comparing this position with
the \HI\ map of Young \& Lo (1996)\markcite{YL96} shows that the claimed
CO detection arises in a large depression in the \HI\ column density.
We detected CO at {\it none} of the three positions, including that observed
by Tacconi \& Young\markcite{TY87}.  The non-detection of CO is in accord 
with the low metallicity of Leo~A and the non-detection of other systems 
with similarly low metallicities.  The rms noise for the spectra we have 
obtained in Leo~A range from 2.7 to 5.5 mK when smoothed to 5.2 \kms\
velocity resolution.  For a resolution of 20.8 \kms\ the range is 1.3 to 
2.6 mK.  In comparison, the detection from Tacconi \& Young\markcite{TY87} 
is 19 mK with a velocity width of 25 \kms.  

{\it Sextans~A---}  The dwarf irregular galaxy Sextans~A was most 
recently observed in CO (prior to our own observations) by Ohta \etal
(1993)\markcite{OTSSN}.  They observed a single position coinciding with
the peak \HI\ column density determined from the map of Skillman \etal
(1988)\markcite{STTwW}, attaining an rms noise in their spectrum of 48
mK for a velocity resolution of 2.6 \kms.  We have observed two positions 
in Sextans~A, the main peak in the \HI\ column density, as well as a 
secondary peak.  These two peaks are positioned on either side of a 
depression in the \HI\ column density that coincides with the center of 
the optical galaxy.  In our observations of the secondary \HI\ peak 
(labeled as position 3 in Table~1) we used the the 
\HI\ hole as the position for the reference beam.  Thus emission at that
location would appear as an absorption feature in the spectrum of Figure~1.
The velocity difference between the gas at these two locations due to the
rotation of the galaxy is large enough ($\sim$ 10 \kms) that an apparent
absorption feature would not cancel out emission at position 3.  The rms
noise in our spectra span the range 3.4 to 5.5 mK at 5.2 \kms\ velocity 
resolution, and 1.3 to 2.4 mK for 20.8 \kms\ resolution.

{\it DDO~210---}  DDO~210, another gas rich dwarf irregular galaxy, is
among those observed in CO by Tacconi \& Young (1987)\markcite{TY87},
who did not detect it.  They give a 2$\sigma$ upper limit on the
integrated CO intensity, I$_{CO}$ of 0.41 K \kms, compared to our
5$\sigma$ upper limit of 0.12 K \kms.  We used the \HI\ maps by Lo,
Sargent, \& Young (1993)\markcite{LSY} to direct our single pointing
observation at the peak of the \HI\ column density.  The Tacconi \&
Young (1987)\markcite{TY87} position falls approximately one beam width
to the east of ours, although still on an area of high \HI\ column
density.  We note that the oxygen abundance we use for DDO~210 
is derived from its absolute magnitude using the luminosity--metallicity
relation of Skillman, Kennicutt \& Hodge (1989).  This is necessary
because it does not have \HII\ regions in which oxygen lines can
be observed (van Zee, Haynes \& Salzer 1997).

{\it DDO~187---}  DDO~187 is also a dwarf irregular, but it has not
been previously observed in CO.  We obtained a spectrum from a single
position in the galaxy, centered on the peak of the \HI\ column density
as determined from the data of Lo, Sargent, \& Young
(1993)\markcite{LSY}.  Smoothing the spectrum to 5.2 \kms, we reach an
rms noise of 2.7 mK, while for 20.8 \kms\ the noise is 1.6 mK.

{\it Pegasus---}  The Pegasus dwarf irregular galaxy has also been observed
in CO by Tacconi \& Young (1987)\markcite{TY87}, who do not detect it,
giving a 2$\sigma$ upper limit on the integrated CO intensity of 0.38
K \kms.  This compares to our 5$\sigma$ upper limit of 0.080 K \kms.
The position observed by Tacconi \& Young (1987)\markcite{TY87} is 
$\sim$ 1\arcmin\ north of the position we have observed, and is on the
edge of the dense region of \HI\ for which our observed position is the 
\HI\ peak.

{\it NGC 5253---} The amorphous galaxy NGC~5253 is a 4 $\sigma$
detection in the smoothed spectrum, with a peak intensity at roughly
400 \kms, the systemic velocity of the \HI\ distribution.  Turner, Beck
\& Hurt (1997) also obtained a detection (14 Jy \kms\ =
6 -- 8$\sigma$) of NGC~5253 at the velocity of the \HI\ using the Owens
Valley Radio Observatory millimeter array.  Using the SEST telescope, 
Wiklind \& Henkel (1989) find an integrated CO intensity of 1.3 K \kms, which 
corresponds to 27.3 Jy \kms\ assuming a gain of 21 Jy/K at 115 GHz.  For 
the 12~m telescope, we adopt a gain of 34 Jy/K (NRAO 12m user's guide)
which yields 24.6$\pm$5.0 Jy \kms, consistent with the SEST value, and 
roughly twice the OVRO interferometer value.  The discrepancy between
single dish and interferometer measurements is to be expected if the 
interferometer resolves out CO emission on large angular scales.  NGC 5253 
was also observed by Jackson et al. \markcite{JSHB}(1989), who did not 
detect it with the NRAO 12-m telescope, with an rms noise of 0.10 K in 
their spectrum, a value slightly higher than in our own spectrum. 

{\it NGC 1569---} NGC~1569 has been classified as a Magellanic irregular. 
Israel \& de Bruyn\markcite{IdB88} (1988) have found evidence that it 
is in a post starburst phase, in which the massive star 
formation has recently ceased.  For this galaxy we find an integrated CO 
intensity of 0.685 $\pm$ 0.104 K \kms.  The CO spectrum shows an absorption 
signature near 0 \kms\ due to Galactic foreground CO emission in the 
reference beam.  Rogstad \etal\markcite{RRW} (1967) obtained an \HI\ 
spectrum towards NGC 1569 which shows \HI\ emission at $\sim$ $-$90 \kms, 
and also detected a narrow emission feature at $\sim$ 0 \kms\ from the Galaxy.  
The two features are well separated so there should not be any contamination 
from the off-beam reference position affecting the line profile.  Israel
Tacconi, \& Baas \markcite{ITB95}(1995) obtained an upper limit with a 
peak T$_R^*$ of 0.024 K.  Young, Gallagher, \& Hunter\markcite{YGH} 
(1984) detected NGC~1569 at I$_{CO}$ = 1.6 $\pm$ 0.3 K \kms, more than 
twice our detected level.  We do not know the reason for this difference, 
but we note that the Israel et al. \markcite{ITB95}(1995) value is consistent 
with our measurement and inconsistent with Young \etal\markcite{YGH} 
(1984).  Greve \etal\markcite{GBJM}(1996) have mapped the CO distribution in 
NGC~1569 in both the 1$ \rightarrow$ 0 and 2$ \rightarrow$ 1 lines.  In these
maps, the CO emission is confined to an area approximately 20\arcsec 
$\times$ 20\arcsec, small enough to be contained in our 55\arcsec beam.

{\it NGC 4214---} NGC~4214 is classified as SBmIII and is one of the 
most luminous of the observed galaxies.  Four positions separated by 
45\arcsec, were observed.  These are designated a,b,d and e.  
CO is detected at 2 of the 4 positions, with an integrated CO 
intensity of 0.900 $\pm$ 0.105 K \kms.  A direct comparison of 
these results to other CO observations of NGC 4214 is problematic 
because of different beam sizes (IRAM; Becker \etal\ 1995) or overlapping 
beams (Thronson et al.\markcite{THTGH} 1988).  Our results are at least 
consistent with the center position 3$\sigma$ detections reported by 
Thronson \etal\ (I$_{CO}$ = 1.0 $\pm$ 0.35 K\kms) and Tacconi \& 
Young\markcite{TY85} (1985; I$_{CO}$ = 0.94 $\pm$ 0.22 K\kms).  
Becker \etal\markcite{BHDW}(1995) have mapped this galaxy in the $^{12}$CO
2$ \rightarrow$ 1 transition, finding the emission to be confined to a 
region roughly thirty arcseconds in diameter.  Our observations cover this
area and should detect all the $^{12}$CO 1$ \rightarrow$ 0 emission. 

{\it Mrk 178---} 
Mrk~178 is a low luminosity dwarf with an abundance of 12+log(O/H) $=$
8.0 (Kobulnicky \& Skillman, in prep.)
Mrk~178 has been observed in CO previously, but has only a high upper limit
(rms noise $=$ 20 mK, Morris \& Lo 1978\markcite{ML78}).  
Our spectrum shows no emission down to a limit of $I_{CO}$$<$0.2 K \kms.  
The rms noise in the final averaged spectra is 2 mK, a factor of 10 
lower than that obtained by Morris \& Lo\markcite{ML78} (1978). 

{\it UGC~4483---}
UGC~4483 is a very low abundance dwarf galaxy (12 +log(O/H) $=$ 7.5) in the 
nearby M81 group (Skillman \etal\ 1994).  Its appearance is 
dominated by a single giant star forming complex.  It has no previous
CO observations in the literature, and no CO emission was detected in 
our spectrum, which has an rms noise of 2 mK. The upper limit on 
I$_{CO}$ is $<$ 0.195 K \kms. 

{\it UM422---}  The \HII\ galaxy UM422 (UGC~6345) is an emission line
galaxy from the sample of Salzer, MacAlpine, \& Boroson 
(1989)\markcite{SMB}, and was
included in the VLA \HI\ survey of \HII\ galaxies of Taylor \etal 
(1995)\markcite{TBGS}.  It is the most distant galaxy in this paper,
and was included in our sample because of its low metal abundance.
This galaxy has not been previous observed in CO.  The rms noise in
our spectrum is 2 mK, with an upper limit on I$_{CO}$ of 0.120 K \kms.

Table~2 gives distances, optical luminosities, oxygen abundances, 
integrated CO intensities and CO ``luminosities'' for galaxies we have 
observed.   L$_{CO}$ is determined using the relation:
L$_{CO}$ = I$_{CO}$ A$_S$, where A$_S$ is the source area (\eg\ Sanders,
Scoville \& Soifer 1991).   

In the case of UM422, the most distant galaxy we observed, the star 
forming region traced by H$\alpha$ emission is approximately the size of 
the telescope beam.  Thus this assumption of extended emission is
adequate, as long as spatially extended star formation can be taken as 
indicating the presence of spatially extended molecular gas.  Without a
knowledge of the true CO spatial distribution, 
our assumption will at least provide reasonable {\it relative} estimates 
for the CO luminosities and H$_2$ masses among galaxies in our sample.  

\placetable{tab2}

\section{Discussion}

\subsection{The Dependence of CO Emission on Metal Abundance}

We have carried out these observations with the goal of better
understanding the relationship of CO emission to metal abundance in 
dwarf galaxies.  Because all of these galaxies (with the exception of 
DDO~210) are currently
experiencing at least some massive star formation, we can infer the
presence of molecular gas.  Even dwarf irregular galaxies with
relatively low rates of massive star formation can be detected in CO,
especially if they are nearby.  To supplement our eleven galaxies we
have searched the literature for low luminosity dwarfs which have been
observed in the $^{12}$CO 1 $\rightarrow$ 0 line.  These previous works
include Morris \& Lo\markcite{ML78} (1978), Rowan--Robinson, Phillips
\& White\markcite{RR80} (1980), Elmegreen, Elmegreen \&
Morris\markcite{Ee80} (1980), Israel \& Burton\markcite{IB86} (1986),
Tacconi \& Young\markcite{TY87} (1987), Thronson \&
Bally\markcite{TB87} (1987), Arnault \etal\markcite{ACCK}(1988), Sage
\etal\markcite{SSHH}(1992), Wilson\markcite{W92} (1992), Hunter \&
Sage\markcite{HS93} (1993), Brinks \& Taylor\markcite{BT} (1995),
Israel, Tacconi \& Baas\markcite{ITB95}(1995), 
Young \etal\markcite{Y95}(1995) and Gondhalekar
\etal\markcite{GJBGB}(1998).  In an effort to keep the galaxy sample
as homogeneous as possible, we selected only those galaxies with M$_B$
$\geq$ -18.  For CO data, we are careful to convert
I$_{CO}$ from the published temperature units (usually $T_A*$ or
$T_{mb}$) into the units used here, $T_R*$.  Since most galaxies have
not been mapped in CO with an interferometer, we use the telescope beam
size at the distance of the galaxy as a consistent first order estimate
the source diameter.  The data we have collected from the literature
are presented in Table~3.  Unfortunately, only a small fraction of
these galaxies have published chemical abundances from optical
spectroscopy.  

\placetable{tab3}

Our primary objective is to study the dependence of CO emission on
ambient metal abundance, independent of variations in galaxy size,
distance, and optical luminosity.   Verter \& Hodge (1995) and Wilson
(1995) use observations of individual molecular clouds in nearby
galaxies to characterize the variation of $X_{CO}$ with metallicity.
Our data do not allow us to determine this conversion factor.
Instead, our approach is to study a larger sample of extremely metal-poor
galaxies using very sensitive CO observations obtained with a telescope
beam that is comparable to the size of the target galaxies.  Since the
galaxies lie at different distances up to 20 Mpc, the telescope beam
samples an ensemble average of many molecular clouds in the target
galaxies.  In most cases, the size of the CO emitting region is not
known.  Thus, the spectra for each pointing represent a mean CO surface
brightness for the nearby resolved galaxies, and a lower limit on the
CO surface brightness for the most distant objects where the emitting
region may be much smaller than the beam.  It is not immediately
apparent which physical properties (e.g., I$_{CO}$, L$_{CO}$, 
M$_{H_2}$) derived from the spectra make for the best analysis.

We first examined the CO luminosity, L$_{CO}$, as a function of metal
abundance as indicated by 12 + log(O/H).  However, because metallicity
correlates strongly with optical luminosity in galaxies of all types
(e.g., Lequeux \etal\ 1979; Skillman, Kennicutt, \& Hodge 1989) we find
that more metal rich, and thus the most luminous, galaxies have larger L$_{CO}$.  This result is not
especially informative.  Larger galaxies contain more matter, and not
surprisingly, they should have more CO as well, even if the
I$_{CO}$/N$_{H_2}$ conversion factor is identical to smaller, more
metal--poor galaxies.  Next, we considered normalizing the CO
luminosity of each galaxy by some fiducial indicator of its mass, such
as optical luminosity or \HI\ mass.  This has the advantage of
producing distance--independent quantities like L$_{CO}$/M$_{HI}$ or
L$_{CO}$/L$_B$.  However, we find that for a given metal abundance, the
scatter in each of these exceeds an order of magnitude.  This scatter 
results, in part, because 
the CO, \HI\, and optical data sample different regions of the galaxy. 
 The \HI\ 
and optical measurements refer to the global properties of a galaxy, 
including material at large radii, while, in all but the most distant 
targets, the CO data represent relatively localized measurements which 
cover only the central starforming regions. Furthermore, extinction and 
the recent star formation history strongly influences the measured optical 
luminosity.  Ideally, such a normalization by optical luminosity or \HI\ 
mass should be made using optical imaging or aperture synthesis \HI\ 
mapping which is spatially matched to the single--dish CO beam.  

We finally decided to use the integrated CO intensity, I$_{CO}$, as
the unit of comparison between galaxies of different metallicity.
I$_{CO}$ is a measure of the mean CO surface brightness (K \kms 
beam$^{-1}$), and is roughly independent of distance as long as the CO
beam is not much larger than the emitting region.  Since I$_{CO}$ 
measures the amount of CO emission per unit beam area, the major problem
with this quantity is that the CO beam subtends larger areas with 
increasing galaxy distance.  For more distant galaxies which are 
smaller than the beam area, the measured quantity represents only a 
lower limit on the CO surface brightness.  In an effort to make a 
robust comparison between galaxies, we exclude from further analysis 
all objects more distant than 10 Mpc.  At 10 Mpc, the 55\arcsec\ FWHM 
beam of the NRAO 12m telescope used in most of these observations subtends 
2.7 kpc.  This is approximately the size of CO emitting central regions 
of Magellanic irregular galaxies such as NGC~4214 (Becker \etal 1995).  CO
detections with 55\arcsec\ resolution in dwarf galaxies significantly
more distant than 10 Mpc will yield probable lower limits on the CO
surface brightness,  I$_{CO}$.   The opposite problem exists for very
nearby galaxies where the telescope beam resolves individual giant
molecular clouds.  The CO data of Israel \etal (1993) in the LMC and
SMC show a larger scatter in I$_{CO}$ which probably reflects the real
brightness variations between molecular clouds centers and inter-cloud
regions.  We include the LMC in the plot for comparison purposes, 
although it violates our absolute magnitude limit.  For the LMC
and SMC we adopt the mean I$_{CO}$ values.  Since Israel \etal (1993) 
chose locations in the LMC and SMC to contain molecular clouds and
star-forming regions, this mean represents an upper limit on the true
mean I$_{CO}$ that would be observed from a distance of 2-5 Mpc which
is typical of the dwarf galaxies under consideration.  The rest of the
galaxies in the sample (except NGC~6822) are more distant than 1 Mpc,
so that these smaller scale brightness variations are smoothed out by
the relatively larger beam area.

In Figure~2 we plot log I$_{CO}$ versus the oxygen abundance (12 + log
(O/H)) for dwarf galaxies less than 10 Mpc away.  Galaxies with CO
detections appear as filled circles, while undetected objects 
appear as upper limits.  We plot I~Zw~36 with an open circle since it
represents a less secure (4$\sigma$) detection (Tacconi \& Young
1987\markcite{TY87}; Young \etal\ 1995) and it has not been
subsequently re-observed.  We show only the brightest position for
NGC~6822 reported by Wilson (1992).  We plot the mean I$_{CO}$
for the LMC and SMC reported by Israel \etal\ (1993).  

Figure~2 reveals a clear dichotomy between systems with 12 + log(O/H) $>$ 8.0
and the very metal-poor systems.  All of the galaxies with CO
detections have higher metallicities; the only one detected below 12 +
log (O/H) = 8.0 is I~Zw~36 (Tacconi \& Young 1987\markcite{TY87}).  All
galaxies with lower metallicities are non-detections with very low
limits.  The non-detections at low metallicities are consistent with a
strong dependence of the CO surface brightness on metallicity.  To test this
visual impression of the data quantitatively, we randomly redistributed
the $x$ and $y$ values of the 19 objects in Figure~2 (I$_{CO}$ and  
12 + log(O/H)) 100,000
times.  In only 97 of those 100,000 tests did all 8 detected objects in
Figure~2 fall above an oxygen abundance of 7.9.  The chance of
obtaining the result randomly is only 0.1\%, strongly suggesting that
metal-poor dwarfs have markedly lower CO surface brightnesses.

\placefigure{fig2}

In previous works there has also been a clear trend for high metallicity
galaxies to have a high CO emission, while most of the low
abundance galaxies were undetected.  Tacconi \& Young\markcite{TY85}
(1985) noted a dependence of L$_{CO}$ on metal abundance, although
their sample of galaxies contained relatively massive, metal-rich
objects and only 1 object with 12 + log (O/H) $<$ 8.5.  They presented
a plot similar to our Figure~2, showing a clear correlation of L$_{CO}$
with O/H for irregular and spiral galaxies, albeit with considerable
scatter.  Gondhalekar \etal\markcite{GJBGB}(1998) also find a similar
result.  Part of this correlation was undoubtedly due to the underlying
luminosity---metal abundance correlation among galaxies ({\it e.g.}
Lequeux \etal\markcite{LPRST}1979, Skillman, Kennicutt, \& Hodge\markcite{SKH}
1989).  Arnault \etal\markcite{ACCK}(1988) made a similar plot using
L$_{CO}$/M$_{HI}$, which shows a correlation between L$_{CO}$/M$_{HI}$
and oxygen abundance.  They include spiral galaxies while we
specifically excluded spiral galaxies from our plot, both because the
physical conditions of the molecular gas are likely to be different
from dwarf galaxies, and because the concept of a global metal
abundance is ill-defined.  Spiral galaxies often show large abundance
gradients (\eg\ Zaritsky, Kennicutt, \& Huchra \markcite{ZKH}1994),
whereas dwarf and irregular galaxies have very uniform abundances
(Pagel \etal\markcite{Pe80}1980; Kobulnicky \& Skillman
\markcite{KS96}1996, \markcite{KS97}1997; Devost, Roy \& Drissen
\markcite{DR}1997).  Sage \etal \markcite{SSHH}(1992) present 
much the same plot (their Fig.~4a) from which they concluded
that there is {\it no} correlation between L$_{CO}$/M$_{HI}$ and metal
abundance.  Our restricted sample includes none of the Sage
\etal galaxies, which are all more distant than 10 Mpc or more luminous
than dwarf galaxies.

Figure~2 should be free from all of these biases due to differing
distances, galaxy sizes, metallicities, and luminosities.  We plot CO
surface brightness rather than luminosity, and we include only dwarf
galaxies which are chemically homogeneous.  We also extend the
metallicity baseline to much lower values of O/H in order to place
stronger constraints on the role of metallicity in determining the CO
surface brightness.  Unfortunately, Figure~2 is relatively sparse
because so few dwarf galaxies have sensitive CO observations and few of
those have accurate metallicity determinations.

To increase the sample size using more of the galaxies
from Table~3, we plot $I_{CO}$ versus L$_{B}$ in Figure 3.  Given the
metallicity -- luminosity relationship for dwarf irregular galaxies,
the luminosity can serve as a metallicity and size indicator.  Mindful
of the historical problems with false CO detections at low signal-to
noise (e.g., Leo A) we further impose the restriction that the CO
detection must be at the 4$\sigma$ level or better.  This excludes
three objects, NGC~3738, DDO~83, and DDO~68, from Table~3.  Labels and
filled circles or large arrows denote galaxies with measured
metallicities which appear in Figure~2.  Filled triangles denote
additional CO detections in objects without measured metallicities.
Small arrows mark additional upper limits for galaxies which have been
observed in CO.  The addition of these 34 galaxies reinforces the
striking trend seen in Figure~2.   Galaxies detected in CO cluster near
log I$_{CO}=0$ and have M$_B$ brighter than $-15.5$.  No galaxies fainter
than M$_B = -15$ are detected, with the exception of the 4$\sigma$ I~Zw~36.
Because I~Zw~36 stands out in this way in Figures~2 and 3, it would
be worthwhile to reobserve it for confirmation of the detection.  The 
CO upper limits of the additional data from the literature do not 
constrain the behavior of I$_{CO}$ at low metallicities as strongly 
as the new observations presented here.

\placefigure{fig3}

Since CO emission is considered a tracer of the molecular gas, it might
be expected from the above result that low abundance galaxies would
also be deficient in molecular gas compared to the amount of atomic
gas.  However, the conversion rate from CO to H$_2$ depends on
abundance, in the sense that the lower the abundance, the higher the
conversion rate.  Therefore the low abundance galaxies will have the
highest conversion rates (e.g., MB88), and thus may not necessarily 
have lower H$_2$
masses.  The best evidence for a metallicity-dependent conversion
factor dependent comes from CO observations of giant molecular clouds
in metal poor systems in nearby galaxies (Verter \& Hodge\markcite{VH}
1995, W95\markcite{W95}).  The new, very sensitive data which we
present here further strengthen their conclusions, and even suggest a
rapid (non-linear) increase in $X_{CO}$ below 12+log(O/H)=8.0.  Spaans
\etal (1998) find such a sharp change in $X_{CO}$ at approximately
this same metallicity in their models of the multi-phase galactic
medium.  Even more sensitive observations with the next generation of
millimeter-wave telescopes may be able to confirm this prediction of a
steep decline in CO surface brightness, and a steep increase in the
I$_{CO}$/H$_2$ conversion factor in very metal-deficient environments.

\section{Summary}

$^{12}$CO $1\rightarrow 0$ observations of 11 galaxies with oxygen
abundances 12 + log (O/H) in the range 8.4 to 7.3 yield the most
sensitive data yet on very metal-deficient galaxies.  The six objects
which have low abundances (12+log(O/H)$<$8.0) are not detected to upper
limits of 0.1 K \kms.  Three of these six have never before been
observed in a CO line, while the others now have much more stringent
upper limits.  For the very low metallicity galaxy Leo~A, we do not
confirm a previously reported detection in CO, but the upper limit
is consistent with an unpublished nondetection by L. Young (1997, 
private communication).

We combine these new observations with data from the literature to form
a sample of dwarf galaxies which all have CO observations and measured
oxygen abundances.  None of the galaxies with 12+log(O/H)$<$7.9 are
detected.  Most of the galaxies with higher metallicities are detected
at a similar CO surface brightness, log I$_{CO}\simeq-0.1$ K \kms.  
These data are consistent
with a strong dependence of the I$_{CO}$/M$_{H_2}~\equiv~X_{CO}$ conversion
factor on ambient metallicity.  The low upper limits on some galaxies,
together with the molecular gas implied by the presence of star formation,
are consistent with hypothesis that the conversion factor is
non-linear, increasing sharply around 1/10 of the solar metallicity
(12+log(O/H)$\sim$8.0).

\acknowledgements

We thank U. Klein for helpful comments on the paper, and C. Wilson for 
commenting on an earlier version of this paper.  We thank the referee
for a thorough review of the paper and useful comments.  We are also 
grateful to the staff of NRAO-Tucson for their assistance with the 
observations.  C.~L.~T. acknowledges support from the Deutsche 
Forschungsgemeinschaft under the framework of the Graduiertenkolleg ``The 
Magellanic System and Other Dwarf Galaxies''.
H.~A.~K. is grateful for assistance from a NASA Graduate Student 
Researchers Program fellowship and from 
\#HF-01094.01-97A awarded by the Space Telescope Science Institute
which is operated by the Association of Universities for Research in
Astronomy, Inc. for NASA under contract NAS 5-26555. 
E.~D.~S. acknowledges support from NASA LTSARP Grant No.  NAGW--3189.

\vfill
\eject

\begin{center}
{\bf Figure Captions}
\end{center}

\figcaption[fig1.ps] { The spectra from the NRAO 12-m telescope.  For each
galaxy the upper spectrum is the original velocity resolution, 2.6 \kms,
while the lower spectrum has been smoothed to 20 \kms.  The horizontal line
at the top of each plot shows the velocity position and full width
at zero max of the \HI\ profile for each galaxy. BPS indicates an observation
in beam switching mode, ABS indicates absolute position switching.  
The galaxies shown are the following: NGC~1569, UGC~4483, Leo~A, Sextans~A,
UM422, Mrk~178, NGC~4214, NGC~5253, DDO~187, DDO~210, and Pegasus.\label{fig1}}

\figcaption[fig2.ps] { Log (I$_{CO}$) versus oxygen abundance, 
12 + log(O/H), for
all known dwarf (M $> -18$) galaxies within 10 Mpc which have CO
observations and measured oxygen abundances.  Filled circles denote
$> 4\sigma$ detections.  Positions of the SMC and LMC are mean values
reported by Israel \etal(1993).  For NGC~4214 and NGC~6822, only the
locations of the peak CO surface brightness are plotted.  The open
circle represents a marginal (4$\sigma$) detection of I~Zw~36  
that has not been confirmed.  All other systems have only upper
limits, including some very low limits on extremely metal-poor systems
reported in this work.  It is striking that no galaxies with
12 + log(O/H) $<$ 7.9 (Z $<$ 0.1 Z$_\odot$) have been detected in CO,
indicating that extremely metal-poor systems have much lower mean CO
surface brightnesses.  This data is consistent with a rise (perhaps
sharply non-linear) in the I$_{CO}$/M$_{H_2}~\equiv~X_{CO}$ conversion factor
at metallicities below 0.1 of the solar value.  \label{fig2}}

\figcaption[fig3.ps] {Log (I$_{CO}$) versus absolute blue magnitude, M$_B$,
for dwarf (M $>-18$) galaxies within 10 Mpc which have CO observations.
A few objects with reported detections at less than the 4$\sigma$ level
appear as upper limits instead (see text).  We plot objects from
Figure~2 using the same filled circles and downward arrows.
Thirty-four additional objects with measured magnitudes but without
measured oxygen abundances appear on this plot compared to Figure~2.
Filled triangles denote the four additional detections, while smaller
arrows without seraphs show the 30 non-detections.  This diagram is
consistent with the clear lack of CO detections at very low
metallicities seen in Figure~2.  However, the upper limits on these
additional data are not especially helpful at constraining the
relationship between CO surface brightness and metallicity.
\label{fig3}}

\vfill \eject

\end{document}


\ptlandscape
\begin{deluxetable}{llccccccccc}
\tablenum{3}
\tablewidth{774pt}
\tablecaption{\Large Data for Galaxies in the Literature \label{tab3}}
\tablehead{\colhead{Galaxy} & \colhead{Galaxy} & \colhead{Distance} & 
\colhead{Reference} & \colhead{M$_B$} &\colhead{Reference} & 
\colhead{Abundance} & \colhead{Reference} & \colhead{I$_{CO}$} & 
\colhead{Reference} & \colhead{L$_{CO}$} \\[.2ex] 
\colhead{Name} & \colhead{Name} & 
\colhead{(Mpc)} & \colhead{}& \colhead{} &\colhead{} & \colhead{12 + log O/H} & 
\colhead{} & \colhead{(K km s$^{-1}$)} & \colhead{} & \colhead{(10$^6$ K km s$^{-1}$ pc$^2$)} }
\startdata
NGC1156 & VV531 & 6.4 & [T88] & -17.5 & [T88] & 8.29(0.20) & [T80] & $<$1.2 & [HS93]
& $<$2.7 \nl
UGCA372 & Mrk67 & 22.4 & [T88] & -16.0 & [TM81] & 8.21(0.08) & [KS96] & $<$0.62 
& [S92] & $<$2.8 \nl
NGC6822 & DDO 209 & 0.7 & [T88] & -15.6 & [T88] & 8.19(0.07) & [KS96] & $>$1.81(0.25)
& [W92] & 0.05 (0.01) \nl
IIZw40 & UGCA116 & 10.3 & [T88] & -17.2 & [T88] & 8.06(0.02) & [KeUP] & 0.46(0.10)
& [Ye95] & 1.82 (0.40) \nl
IIZw70 & Mrk829 & 23.1 & [T88] & -17.1 & [T88] & 8.06(0.08) & [KS96] & $<$0.38 
& [Ye95] & $<$7.6 \nl
UM439 & UGC06578 & 14.7 & [vZ95] & -15.8 & [SMB] & 8.05(0.02) & [KS95] & $<$0.48
& [S92] & $<$5.79 \nl
IZw123 & Mrk487 & 15.0 & [T88] & -16.1 & [T88] & 8.02(0.09) & [KS96] & $<$1.06 
& [Ae88] & $<$2.13 \nl
SMC & \nodata & 0.06 & [BW97] & -16.3 & [BT88] & 7.98(0.02) & [Pe78] & 
1.0(0.5) & [Ie93] &  5200(2600) \nl
UM462 & Mrk1307 & 13.9 & [TBGS] & -15.7 & [S92] & 7.96(0.02) & [KS96] & $<$0.28
& [BT] & $<$3.02 \nl
IZw36 & UGCA281 & 4.7 & [T88] & -13.9 & [T88] & 7.93(0.07) & [VT83] 
& 0.45(0.10) & [Ye95] & 0.37 (0.08) \nl
NGC2366 & DDO 42 & 2.9 & [T88]& -16.7 & [T88] & 7.89(0.02) & [GDe94] & $<$0.60 
& [HS93]& $<$0.28  \nl
UGC3974 & DDO 47 & 2.1 & [T88] & -13.2 & [T88] & 7.85(0.01) & [SKH] & $<$0.46 
& [Ye95] & $<$0.076 \nl
UM461 & \nodata & 11.7 & [S92] & -14.1 & [S92] & 7.76(0.02) & [KS96] & $<$0.78
& [S92] & $<$5.96 \nl 
NGC4789A & DDO 154 & 4.0 & [T88] & -14.1 & [T88] & 7.67(0.06) & [vZ97] & $<$2.6 
& [ML78] & $<$2.32 \nl
GR8 & DDO 155 & 1.7 & [T88] & -11.6 & [T88] & 7.47(0.02) & [SKH] & $<$0.09 & [VH95]
& $<$0.015 \nl
IZw18 & Mrk116 & 14.3 & [T88] & -14.8 & [TM81] & 7.21(0.05) & [SK93] & $<$0.27 & [Ge98]
& $<$1.26 \nl
Arp4 & DDO 14 & 19.8 & [T88] & -17.9 & [T88] & \nodata & \nodata & $<$1.38 & [IB86]
& $<$30.1 \nl
NGC3353 & Haro3 & 16.8 & [T88] & -17.9 & [T88] & \nodata & \nodata & 1.46(0.21) & [S92]
& 3.68 0.53) \nl
UM549 & \nodata & 78.7 & [SMB] & -17.9 & [SMB] & \nodata & \nodata & $<$0.27 & [Ge98]
& $<$37.8 \nl
UGC12632 & DDO 217 & 9.2 & [T88] & -17.7 & [T88] & \nodata & \nodata & $<$0.99 & [IB86]
& $<$4.68  \nl
NGC2537 & Mrk86 & 9.0 & [T88] & -17.6 & [T88] & \nodata & \nodata & 0.83(0.12) & [S92]
& 0.60(0.09) \nl
UGC10310 & DDO 204 & 15.8 & [T88] & -17.5 & [T88] & \nodata & \nodata & $<$3.2 & [ML78]
& $<$44.6 \tablebreak
UM471 & \nodata & 146.6 & [SMB] & -17.5 & [SMB] & \nodata & \nodata & $<$0.46 & [Ge98]
& $<$ 220 \nl
NGC4605 & UGC07831 & 4.0 & [T88] & -17.4 & [T88] & \nodata & \nodata & 1.3(0.3) & [TB87] &
3.39 (0.78)  \nl
HoII & DDO 50 & 4.5 & [T88] & -17.3 & [T88] & \nodata & \nodata & $<$0.46 & [Ye95]
& $<0.35$ \nl
UM465 & UGC6877 & 15.4 & [T88] & -17.3 & [T88] & \nodata & \nodata & 0.63(0.18) & [S92]
&8.34(2.38)  \nl
UGC05478 & DDO 73 & 23.4 & [T88] & -17.2 & [T88] & \nodata & \nodata & $<$0.30 & [Ie95]
& $<$9.17  \nl
NGC4670 & Haro9 & 11.0 & [T88] & -17.2 & [T88] & \nodata & \nodata & $<$2.62 & [Ae88]
& $<$2.83  \nl
UM286 & \nodata & 21.6 & [T88] & -17.0 & [T88] & \nodata & \nodata & $<$0.73 & [Ge98]
& $<$7.70 \nl
UM334 & \nodata & 69.5 & [SMB] & -17.0 & [SMB] & \nodata & \nodata & $<$0.36 & [Ge98]
& $<$39.3 \nl
UM454 & \nodata & 50.1 & [SMB] & -17.0 & [SMB] & \nodata & \nodata & $<$0.73 & [Ge98] 
& $<$41.4 \nl
NGC4144 & UGC7151 & 4.1 & [T88] & -16.9 & [T88] & \nodata & \nodata & $<$4.0 & [RR80]
& $<$3.75 \nl
UM351 & \nodata & 104.2 & [SMB] & -16.9 & [SMB] & \nodata & \nodata & $<$0.27 & [Ge98]
& $<$66.3 \nl
NGC4523 & DDO 135 & 16.8 & [T88] & -16.8 & [T88] & \nodata & \nodata & $<$0.57 & [Ye95]
&$<$6.01 \nl
UM374 & \nodata & 79.1 & [SMB] & -16.8 & [SMB] & \nodata & \nodata & $<$0.46 & [Ge98]
& $<$ 65.1 \nl
NGC7077 & Mrk900 & 13.3 & [T88] & -16.7 & [T88] & \nodata & \nodata & 0.50(0.09) & [S92]
& 0.79(0.14) \nl
UGCA441 & Mrk328 & 19.6 & [T88] & -16.6 & [T88] & \nodata & \nodata & $<$1.04 & [S92]
& $<$3.57 \nl
UM483 & Mrk1313 & 30.8 & [TBGS] & -16.5 & [SMB] & \nodata & \nodata & $<$0.73 & [Ge98]
& $<$15.7 \nl
IC2574 & DDO 81 & 2.7 & [T88] & -16.4 & [T88] & \nodata & \nodata & $<$1.8 & [Ee80]
& $<$0.73 \nl
NGC1560 & UGC03060 & 3.0 & [T88] & -16.4 & [T88] & \nodata & \nodata & $<$4.0 & [RR80]
& $<$2.01 \nl
UM491 & \nodata & 26.3 & [TBGS] & -16.3 & [SMB] & \nodata & \nodata & $<$0.73 & [Ge98]
& $<$11.4 \nl
UM456 & \nodata & 23.3 & [TBGS] & -16.1 & [S92] & \nodata & \nodata & $<$0.64 & [Ge98]
& $<$7.86 \nl
NGC4707 & DDO 150 & 8.0 & [T88] & -16.1 & [T88] & \nodata & \nodata & $<$0.42 & [Ie95]
& $<$1.50  \nl
NGC3077 & UGC05398 & 2.1 & [T88] & -16.1 & [T88] & \nodata & \nodata & 1.59(0.26) & [Ye95]
& 0.26(0.04) \nl
NGC3738 & UGC06565 & 4.3 & [T88] & -16.1 & [T88] & \nodata & \nodata & 0.37(0.10) & [Ye95]
& 0.26(0.07) \nl
NGC2976 & UGC05221 & 2.1 & [T88] & -16.1 & [T88] & \nodata & \nodata & 1.2(0.2) & [TB87] &
0.86 (0.14) \nl
NGC3274 & UGC05721 & 5.9 & [T88] & -16.0 & [T88] & \nodata & \nodata & $<$ 0.1 & [TB87] &
$<$ 0.6 \nl
UM323 & \nodata & 25.5 & [TBGS] & -15.9 & [SMB] & \nodata & \nodata & $<$0.55 & [Ge98]
& $<$8.09 \nl
UM452 & \nodata & 19.2 & [TBGS] & -15.9 & [SMB] & \nodata & \nodata & $<$0.64 & [Ge98]
& $<$5.34 \tablebreak
UGC03966 & DDO 46 & 7.4 & [T88] & -15.5 & [T88] & \nodata & \nodata & $<$4.0 & [RR80]
& $<$12.2 \nl
UGC08320 & DDO 168 & 3.6 & [T88] & -15.5 & [T88] & \nodata & \nodata & $<$4.0 & [RR80]
& $<$2.89 \nl
UM533 & UGC08105 & 10.4 & [S92] & -15.3 & [S92] & \nodata & \nodata & $<$0.87 & [S92]
& $<$5.25  \nl
UGC09405 & DDO194 & 5.7 & [T88] & -15.2 & [MI] & \nodata & \nodata & $<$4.0 & [RR80]
& $<$7.26  \nl
HoI & DDO 63 & 4.4 & [T88] & -15.2 & [T88] & \nodata & \nodata & $<$4.0 & [RR80]
& $<$4.32  \nl
UGC05272 & DDO 64 & 6.5 & [T88] & -15.1 & [T88] & \nodata & \nodata & $<$0.75 & [Ie95] 
& $<$1.77  \nl
UGC02014 & DDO 22 & 9.9 & [T88] & -14.9 & [MI] & \nodata & \nodata & $<$0.95 & [IB86]
& $<$5.20  \nl
UGC05764 & DDO 83 & 6.9 & [T88] & -14.9 & [T88] & \nodata & \nodata & 0.57(0.20) & [Ie95]
& 1.52(0.53) \nl
UGC07698 & DDO 133 & 3.8 & [T88] & -14.9 & [T88] & \nodata & \nodata & $<$4.0 & [RR80]
& $<$3.23 \nl
UGC03860 & DDO 43 & 7.2 & [T88] & -14.6 & [T88] & \nodata & \nodata & $<$0.32 & [Ie95]
& $<$0.93  \nl
UGC07577 & DDO 125 & 3.0 & [T88] & -14.6 & [T88] & \nodata & \nodata & $<$4.0 & [RR80]
& $<$2.01 \nl
VIIZw499 & DDO 165 & 2.8 & [T88] & -14.6 & [T88] & \nodata & \nodata & $<$4.0 & [RR80]
& $<$1.75  \nl
IZw87 & DDO 190 & 3.7 & [T88] & -14.5 & [T88] & \nodata & \nodata & $<$0.46 & [Ie95]
& $<$0.35 \nl 
UGC05340 & DDO 68 & 5.9 & [T88] & -14.4 & [T88] & \nodata & \nodata & 0.30(0.15) & [Ie95]
& 0.58(0.29) \nl
UGC06817 & DDO 99 & 3.1 & [T88] & -14.3 & [T88] & \nodata & \nodata & $<$4.0 & [RR80]
& $<$2.15 \nl
IC1574 & DDO 226 & 4.5 & [T88] & -14.3 & [T88] & \nodata & \nodata & $<$4.0 & [RR80]
& $<$4.52 \nl
UGC06900 & DDO 101 & 5.9 & [T88] & -14.3 & [T88] & \nodata & \nodata & $<$0.75 & [Ie95]
& $<$1.46 \nl
UGC04426& DDO 52 & 6.3 & [T88] & -14.2 & [T88] & \nodata & \nodata & $<$0.63 & [Ie95]
& $<$1.40 \nl
UGC08760 & DDO 183 & 3.3 & [T88] & -13.9 & [T88] & \nodata & \nodata & $<$0.30 & [Ie95]
& $<$0.18 \nl
UGC07559 & DDO 126 & 2.8 & [T88] & -13.2 & [T88] & \nodata & \nodata & $<$4.0 & [RR80]
& $<$1.75 \nl
UGC07599 & DDO 127 & 3.5 & [T88] & -13.0 & [T88] & \nodata & \nodata & $<$4.0 & [RR80]
& $<$2.74 \nl
M81DwA & \nodata & 4.3 & [T88] & -12.3 & [PT] & \nodata & \nodata & $<$0.27 & [Ye95] 
& $<$0.19 \nl
\enddata
\end{deluxetable}